\documentclass[final,5p,times,twocolumn,authoryear]{elsarticle}
\usepackage{lineno}
\usepackage[utf8]{inputenc}
\usepackage{soul}
\usepackage{graphics,graphicx,subcaption}
\usepackage{amsfonts,latexsym,cancel}
\usepackage{ulem}
\usepackage{xcolor}
\usepackage{amssymb}
\usepackage{lipsum}
\usepackage{lineno}
 \usepackage{amsmath}
 \usepackage{multirow}
\usepackage{graphicx}
\usepackage{txfonts}
\usepackage{natbib}
 \usepackage{rotating}
\usepackage[bookmarks=false]{hyperref}

\usepackage{amssymb}
\usepackage{amsmath}

\journal{High Energy Astrophysics}

\begin{document}

\begin{frontmatter}



\title{Dark Matter in White Dwarfs: Implications for Their Structure}


\author[label1]{S\'ilvia P. Nunes}\ead{nunes.silvia@ce.uerj.br}
\author[label2,label3]{Jos\'e D. V. Arba\~nil}
\author[label4]{Juan M. Z. Pretel }
\author[label4]{S\'ergio B. Duarte}

\affiliation[label1]{organization={Instituto de Física Armando Dias Tavares, Universidade do Estado do Rio de Janeiro, Rua São Francisco Xavier 524, 20550-900 Rio de Janeiro, RJ, Brazil}}
\affiliation[label2]{organization={Departamento de Ciencias, Universidad Privada del Norte, Avenida el Sol 461 San Juan de Lurigancho, 15434 Lima,  Peru}}
\affiliation[label3]{organization={Facultad de Ciencias Físicas, Universidad Nacional Mayor de San Marcos, Avenida Venezuela s/n Cercado de Lima, 15081 Lima,  Peru}}
\affiliation[label4]{organization={Centro Brasileiro de Pesquisas Físicas, Rua Dr. Xavier Sigaud, 150 Urca, Rio de Janeiro CEP 22290-180, RJ, Brazil}}

\begin{abstract}
{ White Dwarfs (WDs), the final evolutionary stage of most stars, are frequently modeled considering only a dense plasma matter. However, their potential interaction with dark matter (DM), especially in galactic halos where DM is expected to be prevalent, may lead to significant consequences. This work proposes a novel EoS (EoS) that consistently incorporates both hot dense plasma and cold dark matter (CDM) contributions in hot WDs. The hot dense plasma EoS is extended to include thermal and radiative contributions. At the same time, the CDM component is modeled as a linear fluid, with the coupling constant $\alpha$ determined self-consistently within the star. A smooth phase transition between hot dense plasma and CDM regimes is introduced via a hyperbolic mixing function that depends on local energy density and stellar temperature. Our results show that the inclusion of CDM leads to an increase in the WD radius by approximately $12\%$ and a total mass enhancement of $0.7\%$, compared to standard hot WD models without lattice effects. These results highlight the importance of considering CDM in stellar modeling and suggest that WDs may serve as indirect probes for the astrophysical properties of dark matter.}
\end{abstract}



\begin{keyword}{White Dwarfs, }{Dark Matter Admixture, }{Equation of State}



\end{keyword}

\end{frontmatter}




\section{Introduction} \label{sec:intro}

White Dwarfs (WDs) are remnants of low- to intermediate-mass stars \cite{Iben_1983}, and play a pivotal role in astrophysics, offering valuable insight into stellar evolution. It is estimated that approximately $99\%$ of all stars will eventually evolve into WDs \citep{Kippen_1990}. These objects are distinguished by their degenerate cores, which are stabilized against gravitational collapse by electron degeneracy pressure \citep{Chandrasekhar_1931}. These standard models of WDs typically focus on baryon–lepton matter comprising atoms, almost completely ionized, and fundamental particles.

Although this composition provides a good description of WDs, it overlooks the interactions with another dominant component of galaxies: dark matter. It is now well-established that WDs are primarily distributed in the disks of galaxies such as the Milky Way \citep{Fontaine_2001}. However, there have also been observations of WDs in galactic halos \citep{Kilic_2019,Torres_2021}, the regions where dark matter (DM) is predominantly located as inferred by \cite{Rubin_1970}. This observation raises the intriguing possibility that DM could be accreted into WDs in some way, potentially influencing their properties.

{An investigation into the influence of cold DM (CDM) in WDs was conducted by \cite{Leung_2013}. They modeled WDs as fully degenerate stars hosting an isothermal CDM core, assuming that DM is gravitationally bound and segregated within the central region of the star. This central concentration of DM induces notable modifications in the mass-radius relationship of WDs and may alter their critical mass thresholds. In this article, the authors examine how the presence of a concentrated DM core can significantly impact the structural properties and stability limits of these compact stellar remnants.}

{In \citep{Graham_2018}, the authors investigate the potential of WDs as detectors for heavy DM. They propose that DM particles scattering off the ion constituents within a WD can deposit sufficient energy to locally heat the star by producing high-energy Standard Model particles. The study emphasizes the concept of an ``explosive transit'' of heavy DM through the WD, highlighting the critical role of minimal kinetic energy loss as the DM passes through the star’s nondegenerate outer envelope. This work advances our understanding of how WDs could function as indirect probes of DM properties, particularly regarding its interaction mechanisms and energy deposition in dense stellar environments.}


{The intense gravitational field of WDs, arising from their extreme densities, combined with their typical location in regions of elevated DM density (such as galactic halos) makes them potential regions for the capture and accumulation of DM particles. This interaction could influence the evolutionary trajectory of WDs by altering their cooling rates or generating observable signals associated with dark matter. Although direct observational confirmation remains elusive, we propose a theoretical framework that incorporates the possibility of DM in WD's envelopes, positioning these compact stars as promising laboratories for DM studies.}

{The novel aspect of our work, when compared to existing studies on DM in WDs, lies in the treatment of the stellar core. While some models, such as \cite{Leung_2013}, assume a fully degenerate WD containing dark matter, and others focus on complex interactions within the outer layers \citep{Graham_2018}, our approach employs an equation of state (EoS) that consistently integrates hot dense plasma and cold DM components within hot WDs. Crucially, despite including a non-degenerate DM fraction in the envelope and accounting for thermal and radiative effects in the hot dense plasma EoS, our model preserves the main characteristic of WDs: their cores remain primarily supported by electron degeneracy pressure even at elevated internal temperatures. This strategy enables a realistic exploration of dark matter’s  effects on WD structure and stability, without fundamentally compromising the degeneracy-supported core, thereby maintaining fidelity with established WD physics.}

{In this study, we examine the impact of CDM on the internal structure of WDs by developing a theoretical model that combines a mixed EoS encompassing conventional hot dense plasma (neutrons, protons, electrons, and photons) and a DM component. Motivated by recent observations of WDs in galactic halo regions \citep{Kilic_2019, Torres_2021}, where DM density is expected to be enhanced, our goal is to quantify how CDM presence modifies key structural properties such as mass distribution and stellar radius. To this end, we adopt a thermodynamic formalism that couples the hot dense plasma and DM EoS through a transition function dependent on the local energy density, providing a smooth and continuous interpolation between regimes dominated by hot dense plasma and those influenced by dark matter.}

{This paper is organized as follows: Section \ref{EOS} presents the equations of state for both hot dense plasma and DM components, detailing the mathematical framework employed to couple them, and the stellar structure equations. In Section \ref{results}, the results are reported. In this section, we investigate the effects of DM on the internal structure of WDs, extending the analysis across several stellar masses. Finally, in Section \ref{conclusion} we conclude.}

\section{The EoS and stellar structure equations}\label{EOS}

\subsection{The EoS}

{\it EoS of matter and radiation:} To describe the stellar structure, we adopt an EoS (EoS) for the matter and radiation contributions as proposed by \cite{Timmes_1999} and extended by \cite{Nunes_2021} for the interstellar fluid. In our analysis, we neglect the lattice contribution typically present at lower temperatures, since we are considering high-temperature conditions where the fluid is in a molten state. The chosen EoS incorporates the contributions of nucleons, electrons, and radiation—key components of the interstellar fluid. Nucleons, including protons and neutrons, contribute to the mass and energy density of the fluid. Electrons, although partially degenerate in hot environments, significantly affect the fluid’s pressure. Additionally, radiation, primarily in the form of photons, plays a vital role in both pressure and energy near the stellar surface. By including these components in the EoS, we have 
\begin{eqnarray}\label{pressure}
P_g&=&P_n+P_e+P_\gamma,\\\label{e}
\varepsilon_g&=&\varepsilon_n+\varepsilon_e+\varepsilon_\gamma,
\end{eqnarray}
with the subindices $n$, $e$, and $\gamma$ depicting the nucleons, electrons, and photons contributions, respectively. Regarding the WDs' temperature, the stellar interior is an isothermal core.

{\it DM EoS:} Since the EoS for DM remains unknown, we propose a model that is self-consistent with the structure of WDs. To achieve this, we employ the EoS of DM as follows \citep{Kopp_2018}
\begin{equation}\label{P_DM}
    P_{DM}=\alpha \varepsilon_{DM},
\end{equation}
where $\alpha$ is a constant dependent on the coherence between DM and gas EoS. Here $P_{DM}$ and $\varepsilon_{DM}$ denote the pressure and energy contributions of dark matter. 

To determine $\alpha$, we compare equations (\ref{pressure}) and (\ref{e}) with (\ref{P_DM}). The parameter $\alpha$ is defined at the location where the gas pressure $P_g$ equals the DM pressure $P_{DM}$, and the energy density of the gas $\varepsilon_g $ matches the energy density of DM $\varepsilon_{DM}$. {  The choice of this relation simplifies the comparison with other models such as, for instance, a degenerate Fermi gas. The employ of this relation allows us to (i) regularize the transition between the gas and dark matter components, (ii) keep the problem analytically tractable, and (iii) systematically explore the maximal impact that a self-gravitating dark matter contribution could have on the stellar structure. In this context, the parameter $\alpha$ should be interpreted as an effective parameter encoding the dark matter pressure relative to its energy density. Determining $\alpha$ self-consistently at the point where gas and dark matter pressures coincide provides a controlled prescription for matching the two EoSs, which stabilizes the integration and avoids artificial discontinuities across the interface.}

{\it The mixture phase:} Since we consider the presence of DM in the non-degenerate envelope of WDs, we assume that the star's interior consists of three distinct regions: one composed of electrons, nuclei, and photons, a transition region, and another consisting exclusively of dark matter. To physically describe this mixed phase, we assume that the pressure at a given energy density within the transition region is a function that characterizes the interaction between the two components:
\begin{align}
P(\varepsilon)=f({\varepsilon})P_{DM}+(1-f(\varepsilon))P_{g},
\end{align}
{ where $f(\varepsilon)$ is a function depending on the local density ($\varepsilon$) and the completely degeracy core density ($\varepsilon^\star$) (see \citep{Nunes_2021})}
\begin{equation}
f(\varepsilon)=\frac{1}{2}\left[1+tanh\left(\frac{\bar{\varepsilon}-\varepsilon}{b}\right)\right],
\end{equation}
being 
\begin{equation}
\bar{\varepsilon}=\frac{\varepsilon^\star+\varepsilon_\gamma}{2},
\end{equation}
where $\varepsilon^\star$ is the degeneracy threshold (see \citep{Shapiro,Nunes_2021}, $\varepsilon(P_\gamma)$) and $b=\bar{\epsilon}-\varepsilon_\gamma$. Furthermore, $b$ is recalculated in the code to enforce the conditions $f(r=0)=0$ and $f(r=R)=1$.  {  It is important to note that we chose this function, $f(\varepsilon)$, because it allows us to provide a smooth and continuous transition between the ordinary hot gas and dark matter equations of state. This relation regularizes the crossover, which is necessary to avoid a sharp phase transition, thus preventing any numerical instabilities in the integration that may arise from discontinuities in the equation of state during the phase transition.}

\begin{figure}[!ht] 
\begin{center}
\includegraphics[width=1\linewidth]{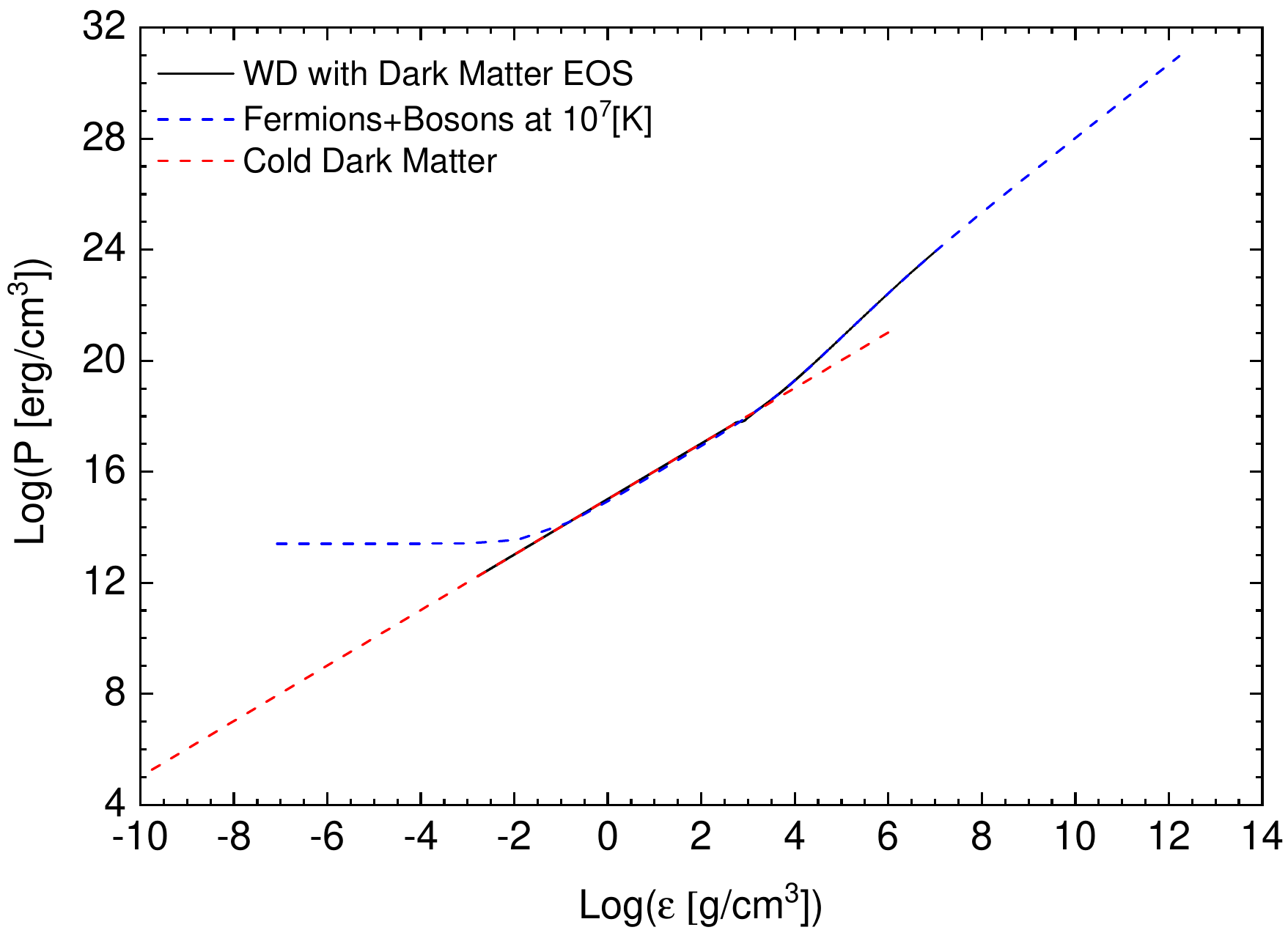}
\caption{Log-log plot of the pressure as a function of energy density for different equations of state (EoS). The blue dashed curve represents a typical WD composed of ions, electrons, and photons at a central temperature of $T = 10^7\,[\rm K]$. The red dashed curve corresponds to CDM. The black curve shows the proposed EoS, which interpolates the standard WD EoS at high densities and the CDM behavior at low densities, incorporating a mixed-phase regime. Temperature effects are also considered, with CDM contributing to the system's cooling, driving it toward $T = 0\,[\rm K]$ at low densities.}\label{fig1}
\includegraphics[width=1\linewidth]{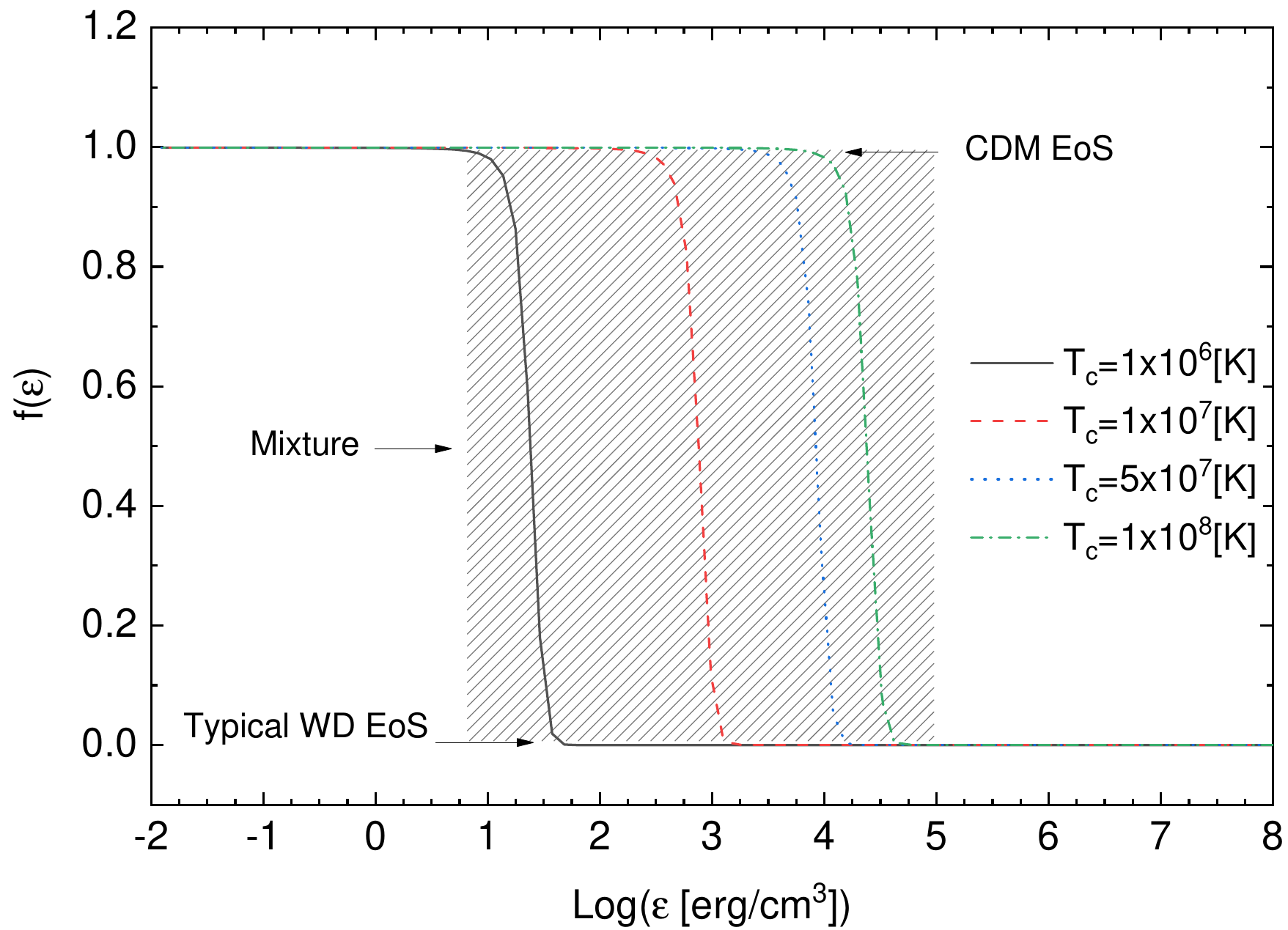}
\caption{Hyperbolic transition function $f(\varepsilon)$ plotted as a function of energy density for four different central temperatures. The function interpolates smoothly between $f(\varepsilon) = 0$, representing a typical WD EoS, and $f(\varepsilon) = 1$, corresponding to a regime dominated by CDM. The transition region reflects a mixed phase where both components coexist.
 } \label{fig2}
\end{center}
\end{figure}

Figure \ref{fig1} presents a comparative analysis of the EoS for WDs in the presence of CDM. The plot displays the relationship between pressure and energy density on a log-log scale, enabling the examination of the EoS behavior over a broad range of values. Three distinct curves are shown: the first (blue dashed line) corresponds to a WD composed of ions, electrons, and photons at a temperature of $T = 10^7\,\mathrm{K}$; the second (red solid line) represents the EoS of CDM; and the third (black line) illustrates the resulting EoS when both components are present, including the limiting cases and the mixed phase.

This figure shows that the EoS proposed (black curve) is consistent with the standard EoS of a typical WD for energy densities above the degeneracy threshold ($\rho_{*}$ following \citep{Shapiro}. Additionally, for densities below the characteristic density associated with the radiative pressure ($\varepsilon(P_{rad})$), the EoS follows that of cold dark matter. In the intermediate regime, our proposed EoS represents a mixture of both components. Beyond pressure and energy density, it is important to emphasize the role of temperature in this formulation. Cold DM acts as a cooling the overall stellar composition, leading the system toward a $T = 0\,\mathrm{K}$ limit at low densities.

In Figure \ref{fig2}, we present the transition function $f(\varepsilon)$ as a function of energy density for different central temperatures: $1 \times 10^6$, $1 \times 10^7$, $5 \times 10^7$, and $1 \times 10^8\ \mathrm{K}$. This function smoothly interpolates between two asymptotic regimes: $f(\varepsilon) = 0$, corresponding to a pure WD EoS, and $f(\varepsilon) = 1$, corresponding to a regime dominated by CDM. The hatched region highlights the intermediate density range where both EoSs contribute, representing a mixed-phase regime. This construction ensures a continuous and differentiable transition between the two physical limits, allowing for a consistent modeling of hybrid stellar structures composed of both hot dense plasma and dark matter.

\subsection{The Stellar Structure Equations}

{To model static, spherically symmetric white dwarfs, we adopt the following spacetime metric expressed in Schwarzschild-like coordinates 
}\begin{equation}\label{metric}
ds^2 = -e^{\nu(r)}dt^2 + e^{\lambda(r)}dr^2 + r^2 d\theta^2 + r^2 \sin^2\theta, d\phi^2,
\end{equation}
{where the metric potentials $\nu(r)$ and $\lambda(r)$ are functions solely of the radial coordinate $r$, reflecting the static and isotropic nature of them.}

{Employing Einstein's field equations for this geometry and assuming a perfect fluid energy-momentum tensor, we derive the Tolman–Oppenheimer–Volkoff (TOV) equations, which govern the hydrostatic equilibrium of relativistic stellar objects \citep{Tolman_1939, OV_1939}}
\begin{align}
\frac{dM}{dr} &= 4\pi r^2 \varepsilon(r), \label{DM}\\
\frac{dP}{dr} &= -(\varepsilon + P)\left( \frac{m(r)}{r^2} + 4\pi r P \right)e^{\lambda(r)}, \label{dp}
\end{align}
{the metric coefficient } $e^{\lambda(r)}$ is determined as
\begin{equation}\label{eq_lambda}
e^{\lambda(r)} = \left(1 - \frac{2m(r)}{r}\right)^{-1}.
\end{equation}

{To obtain stellar equilibrium configurations, we numerically integrate equations \eqref{DM} and \eqref{dp}, starting at the center of the star ($r=0$) with initial conditions $m(0)=0$, $P(0)=P_c$, $\varepsilon(0)=\varepsilon_c$ and $T(0)=T_c$. Integration proceeds outward until the surface is reached at $P(R)=0$.}

\section{Numerical results}\label{results}

\subsection{CDM in white dwarfs}

\begin{figure}[!ht] 
\begin{center}
\includegraphics[width=1\linewidth]{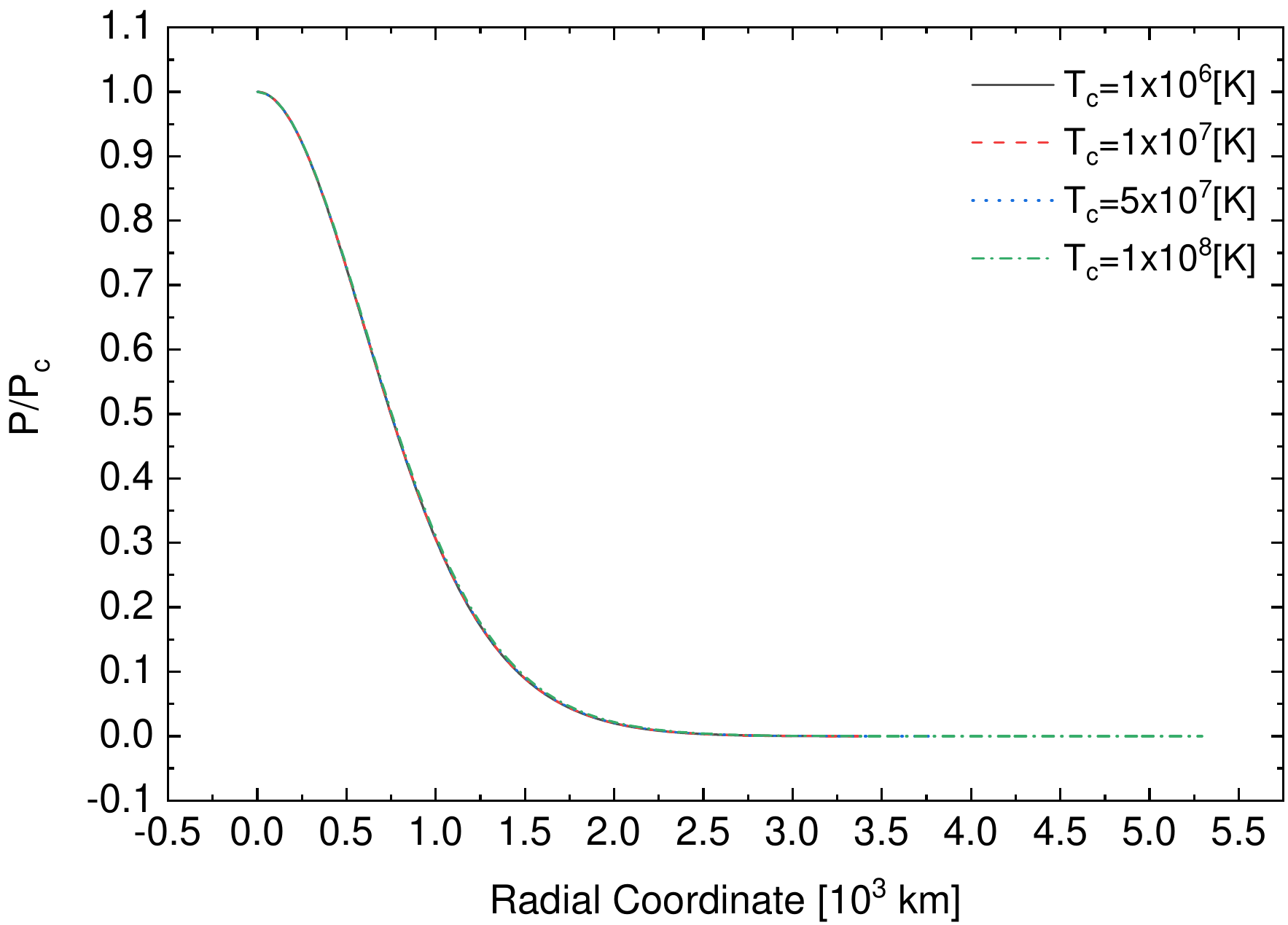}
\includegraphics[width=1\linewidth]{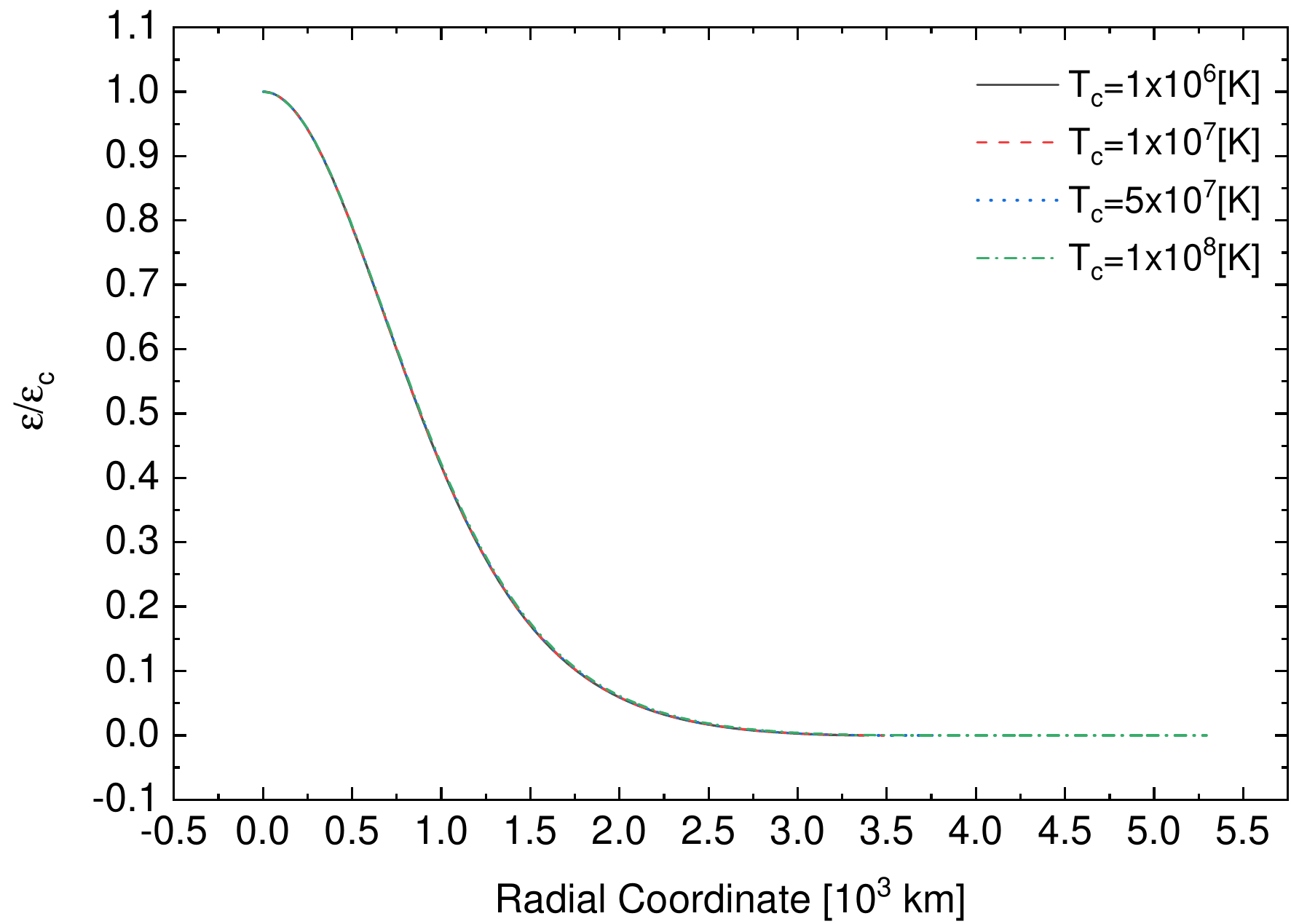}
\includegraphics[width=1\linewidth]{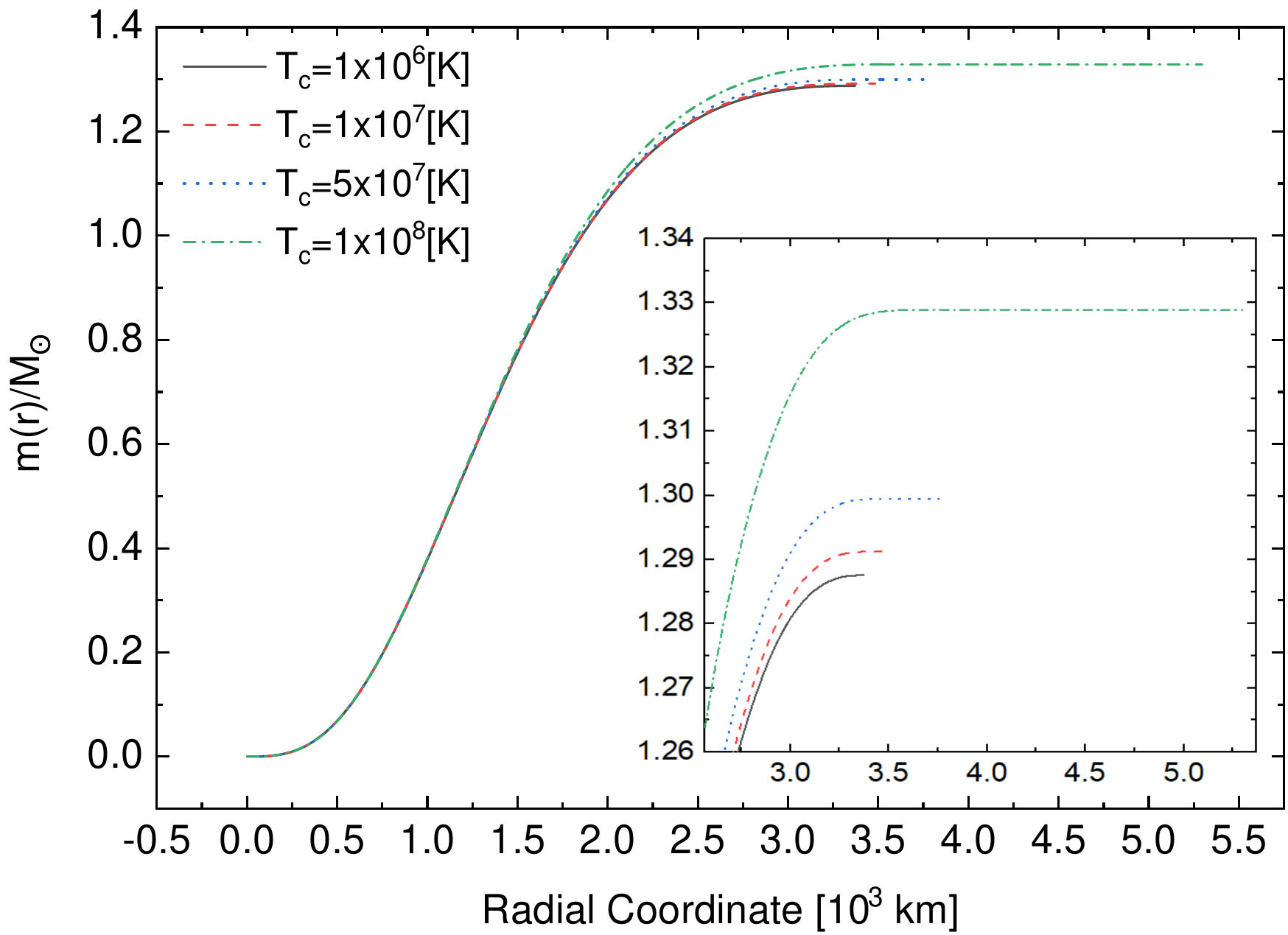}
\caption{{The pressure, energy density, and mass in their normalized form against the radial coordinate are respectively presented on the top, middle, and bottom panels for four different central temperatures. The central pressure and central energy density of normalization are respectively $P_c=9.6\times10^{25} \,[\rm erg/cm^3]$ and $\varepsilon_c=3.0\times10^8\,[\rm g/cm^3]$.}}\label{figb}
\end{center}
\end{figure}

{In Fig. \ref{figb}, the behavior of the central pressure, the energy density, and the mass with the radial coordinate for different values of central temperatures is shown. The central pressure and central energy density considered are $P_c=9.6\times10^{25}\,[\rm erg/cm^3]$ and $\varepsilon_c=3.0\times10^8\,[\rm g/cm^3]$, respectively. In all cases, we can see that the star's pressure and density decrease monotonically with increasing radial coordinate. In turn, the mass increases monotonically with the radial coordinate. From the figure, it can also be observed that pressure, density, and mass are affected by central temperature. In these figures, we can see that the most notable effect is found in the increase in the star's radius.}


\subsection{Several WDs structure}

\begin{figure}[!ht] 
\begin{center}
\includegraphics[width=1\linewidth]{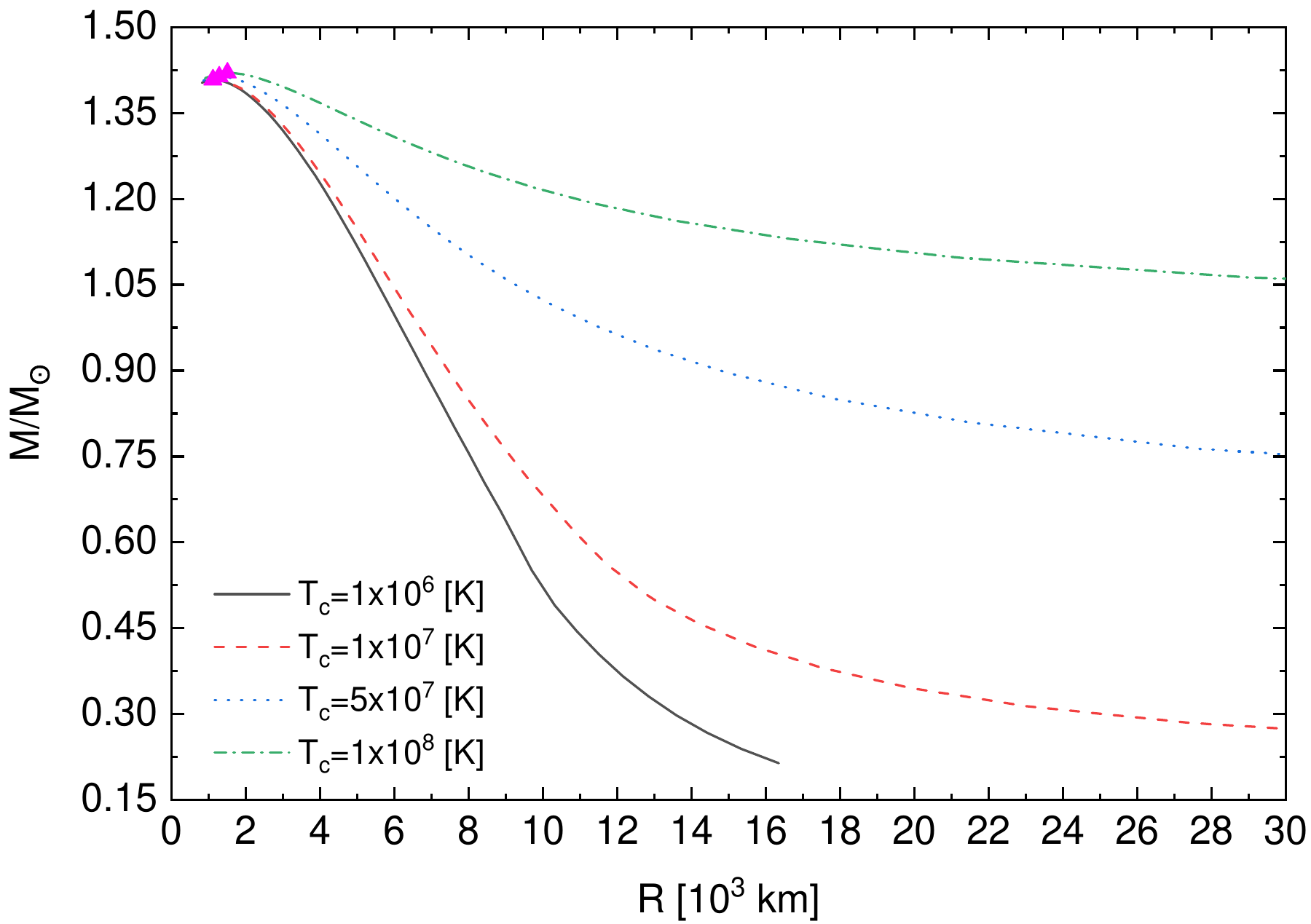}
\caption{Mass as a function of radius for four different central temperatures. Pink triangles indicate the maximum mass configurations, corresponding to the stability limit for each case}\label{fig4}
\includegraphics[width=1\linewidth]{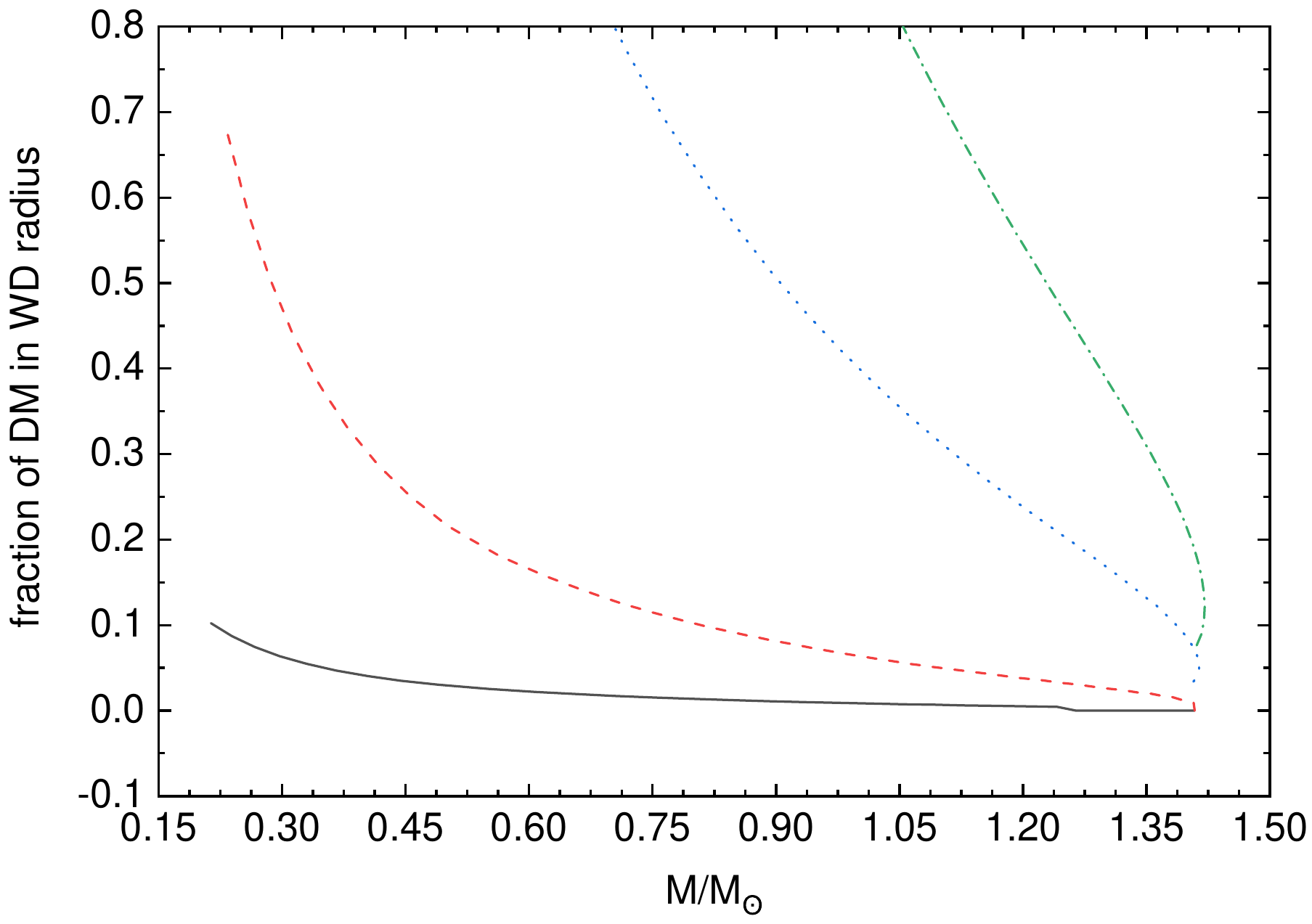}
\caption{The fraction of DM in the WD radius as a function of stellar mass.}\label{fig5}
\end{center}
\end{figure}

In Figure~\ref{fig4}, we present the mass--radius relations for stellar models computed with different central temperatures: \(1 \times 10^6\), \(1 \times 10^7\), \(5 \times 10^7\), and \(1 \times 10^8\,[\rm K]\). Each curve corresponds to a sequence of equilibrium configurations for a fixed central temperature and varying central energy density. The pink triangles denote the turning points in each sequence, which are associated with the onset of dynamical instability against small radial perturbations, typically identified by the condition $DM/d\rho_c=0$. Additionally, in the figure, one observes that for a fixed stellar mass, the stellar radius increases as the central temperature rises. This occurs because higher temperatures enhance the contribution of the thermal pressure, effectively countering gravity, resulting in more extended stellar structures. This behavior is consistent with the results found for typical finite-temperature WDs, as reported by \citet{Nunes_2021}.

As the influence of cold DM (CDM) is more pronounced on the stellar radius than on the stellar mass, we calculate the fractional stellar radius at which CDM begins to mix within the star. This quantity is shown in Figure \ref{fig5}, using the same central temperature values as in Figure \ref{fig4}.
We observe that more massive stars allow a smaller amount of CDM. This behavior arises because, in high-mass models, the stellar core occupies a larger fraction of the star, leaving less room for CDM to be gravitationally confined without disrupting hydrostatic equilibrium. Conversely, as the stellar mass decreases, the star can support a larger fraction of CDM. This is due to the relatively smaller and less dense core, which enables CDM to be accommodated in the outer regions without destabilizing the configuration.

An additional noteworthy trend is that, for a fixed stellar mass, higher central temperatures allow a larger CDM fraction in the stellar envelope. This is because higher temperatures increase the threshold energy density required to break degeneracy pressure, allowing more DM to accumulate before degeneracy is suppressed.

The primary observable parameters for WDs are surface gravity ($Log(g)$) and  effective temperature ($T_{eff}$). To validate our models against observational data, we compare our results with those reported in \cite{Torres_2021} and \cite{Kilic_2019}. These studies focus on WDs located near the Galactic halo, a region where cold DM (CDM) is expected to be more abundant. For consistency, the stellar luminosity of our models is estimated following the prescriptions in \cite{Shapiro} and \cite{Althaus_2010}.

\begin{figure}[!ht] 
\begin{center}
\includegraphics[width=1\linewidth]{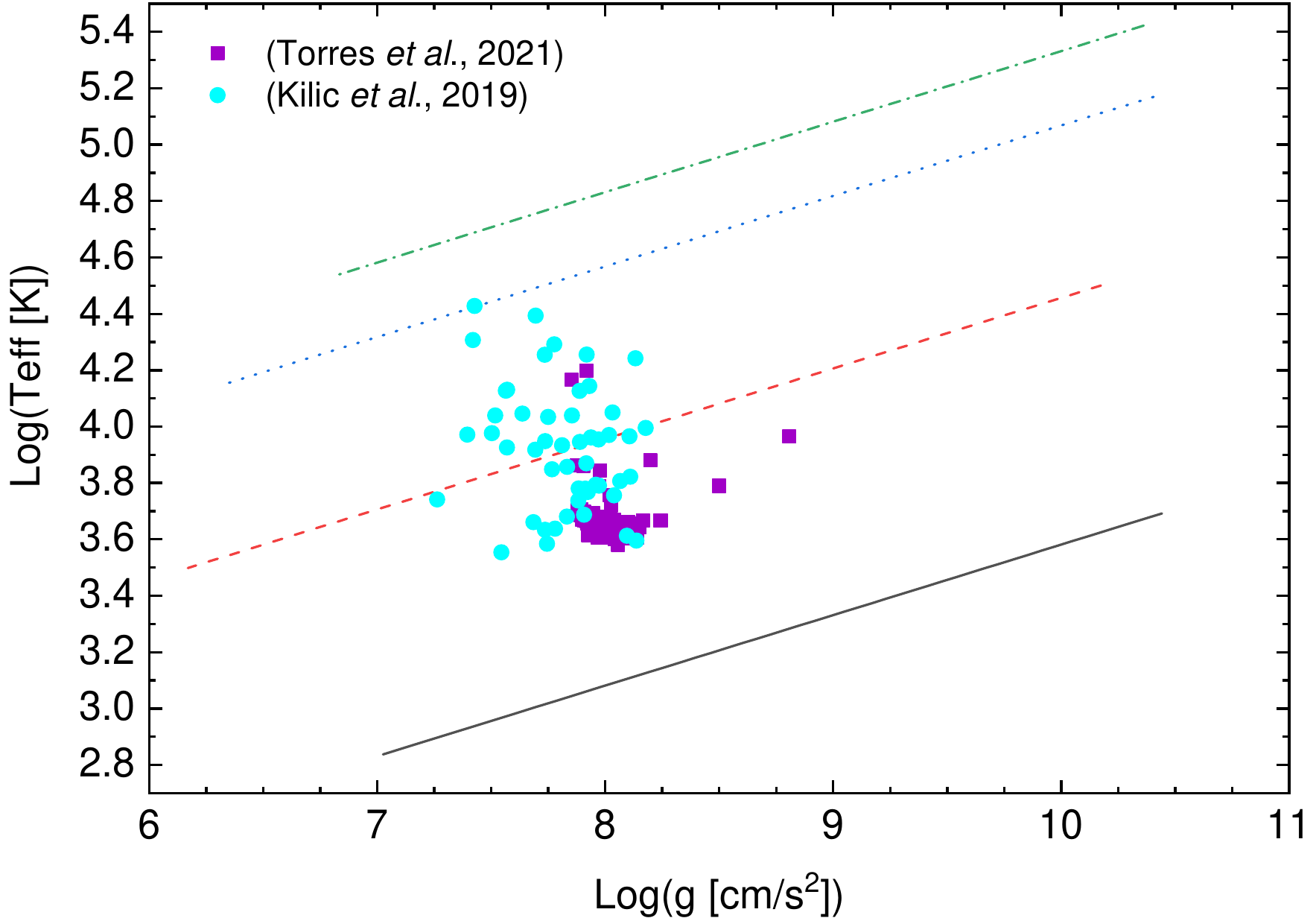}
\caption{The effective temperature against the superficial gravity for different central temperatures. Observational data obtained from the catalogs in \citet{Torres_2021} and \citet{Kilic_2019} are, respectively, marked by purple squares and cyan circles.} \label{fig6}
\end{center}
\end{figure}

Using the computed values of $Log(g)$ and $T_{eff}$, we construct Figure \ref{fig6}, considering the same range of central temperatures as in previous analyses. Observational data from \citet{Torres_2021} are represented as purple squares, while the cyan circles represent the sample from \citet{Kilic_2019}. The theoretical predictions from our models exhibit strong agreement with observations, suggesting that our treatment of CDM aDMixture remains consistent with real stellar populations. This consistency supports the plausibility of DM aDMixture in WDs, particularly in environments with higher CDM density, such as the Galactic halo. Furthermore, the sensitivity of the observable surface gravity to both the internal structure and CDM content offers a promising indirect probe to constrain the presence of DM in compact stellar remnants. Another important behavior to remark is that, according to our model agreement to observational data, the stars remain in the range of $1\times 10^6 \rm [K] \leq T_c \leq 5\times10^7\rm [K]$

It is important to note that we use the typical luminosity of WDs to estimate observable parameters, even in the presence of cold DM (CDM) aDMixture. This approach is justified since, within the regime considered, CDM acts solely as an additional gravitational component in the stellar structure, without significant thermal interactions with hot dense plasma. As CDM does not contribute to radiation emission or directly affect the internal energy transport mechanisms, the classical Stefan-Boltzmann relation remains valid for the visible hot dense plasma component. Therefore, we assume that the luminosity of CDM-aDMixed stars can be described in the same way as for standard WDs, as reported in \cite{Shapiro, Althaus_2010}. This approximation is consistent with the hypothesis that cold DM is gravitationally coupled but thermally decoupled from ordinary matter.

\section{Conclusion}\label{conclusion}

In this work, we developed a new EoS tailored to model WDs with envelopes composed of hot dense plasma and a non-degenerate component of CDM. Our hybrid EoS introduces a continuous interpolation between the pressure contributions of standard WD matter (electrons, ions, and photons) and a cold gas representing CDM, controlled by the dimensionless density parameter $\alpha = P_{\text{DM}}/\varepsilon_{\text{DM}}$. By incorporating a hybrid EoS with a smooth transition between hot dense plasma and dark-matter components, we could model the thermodynamic behavior of a composite stellar fluid composed of electrons, ions, photons, and CDM. Our results demonstrate that, even for modest amounts of dark matter, structural changes emerge in the stellar configuration, most notably, an increase in the radius of the WD and a slight mass enhancement.

The results presented in this work indicate that the CDM contribution in the stellar envelope can lead to measurable deviations in the mass-radius relations. Notably, a non-negligible DM fraction in the outer layers acts {  impacting the matter’s overall thermal profile}. {  It is important to note that the CDM component is thermally decoupled from baryonic matter and does not directly contribute to radiation or internal energy transport; the thermal profile effect refers only to indirect modifications of the temperature gradient caused by the presence of DM.} A consequence of this modeling is that the presence of CDM in the outer layers can induce sufficient cooling to naturally enforce a boundary condition of vanishing temperature at the stellar surface, $T(R)=0$.  This allows us to compare our results with those obtained for WDs near the Galactic halo, assessing whether our proposed model is consistent with current observations in \citep{Kilic_2019, Torres_2021}. Such agreement reinforces the plausibility of our model and highlights its potential to contribute meaningfully to our understanding of white dwarf cooling in non-standard astrophysical environments.

Moreover, our results point to some astrophysical implications. As CDM modifies WD envelope thermal transfer, it could impact age estimates of stellar populations, particularly in old, low-luminosity systems. Additionally, WD with anomalous cooling behaviors may emerge as indirect probes of the galactic DM halo, especially if targeted by future infrared and astrometric surveys such as Euclid and Vera Rubin Observatory.

It is important to mention that our approach contrasts with that of \citet{Leung_2013}, who studied DM effects in the context of fully degenerate WDs with an isothermal CDM core. In their model, DM is gravitationally bound and segregated into the core, leading to more pronounced effects on the mass-radius relation and potential critical mass thresholds (turning-point of stability). {  In particular, adding mass to a dense core enhances the gravitational pull, compressing the star and decreasing its radius, whereas in our model, placing dark matter in the less dense envelope leads to a moderate increase in the stellar radius, ensuring the preservation of hydrostatic equilibrium.} In contrast, our work assumes a non-degenerate DM component in the envelope, resulting in more subtle but qualitatively distinct modifications to the stellar structure. This complementary perspective highlights the importance of exploring a range of DM distributions and interactions in WD models.


\section*{Acknowledgments}
{  We thank the anonymous referee for their valuable suggestions
and comments, which have helped improve this work}. JDVA thank the Universidad Privada del Norte and Universidad Nacional Mayor de San Marcos for the financial support - RR No.$\,005753$-$2021$-R$/$UNMSM under the project number B$21131781$. JMZP acknowledges support from ``Fundação Carlos Chagas Filho de Amparo à Pesquisa do Estado do Rio de Janeiro'' -- FAPERJ, Process SEI-260003/000308/2024. SBD thanks CNPq, Brazil for partial financial support. This work has been done as a part of the Project INCT-Física Nuclear e Aplicações, Brazil , Project number 464898/2014-5..  
\bibliographystyle{apalike}  

\bibliography{bibliography.bib} 


\end{document}